\documentclass[AMA,Times1COL]{WileyNJDv5} 

\articletype{Article Type}%

\received{Date Month Year}
\revised{Date Month Year}
\accepted{Date Month Year}
\journal{Journal}
\volume{00}
\copyyear{2023}
\startpage{1}

\usepackage{graphicx}
\usepackage{textcomp}
\usepackage{xcolor}
\usepackage{amsmath, calc}
\usepackage{tcolorbox}
\usepackage{multirow}
\usepackage{url}
\usepackage{hyperref}
\usepackage{pgfplots}
\pgfplotsset{compat=1.18}
\usepackage{placeins}
\usepackage{fontawesome}
\usepackage{comment}
\usepackage{tabularx}
\usepackage{booktabs}
\usepackage{framed}
\usepackage{fancybox}
\usepackage{makecell}
\usepackage{csquotes}
\usepackage{orcidlink}
\usepackage{multicol}
\usepackage{colortbl}
\usepackage{caption}

\begin{document}

\title{Exploring Indicators of Developers’ Sentiment Perceptions in Student Software Projects}

\author[1]{Martin Obaidi}
\author[1]{Marc Herrmann}
\author[1]{Jendrik Martensen}
\author[2]{Jil Klünder}
\author[1]{Kurt Schneider}


\address[1]{\orgdiv{Leibniz Universität Hannover}, \orgname{Software Engineering Group}, \orgaddress{\city{Hannover}, \country{Germany}}}
\address[2]{\orgdiv{University of Applied Sciences | FHDW Hannover}, \orgaddress{\city{Hannover}, \country{Germany}}}

\corres{Corresponding author: Martin Obaidi, \email{martin.obaidi@inf.uni-hannover.de}}



\abstract[Abstract]
{Communication is a crucial social factor in the success of software projects, as positively or negatively perceived statements can influence how recipients feel and affect team collaboration through emotional contagion. Whether a developer perceives a written message as positive, negative, or neutral is likely shaped by multiple factors.

In this paper, we investigate how mood traits and states, life circumstances, project phases, and group dynamics relate to the perception of text-based messages in software development. We conducted a four-round survey study with 81 students in team-based software projects. Across rounds, participants reported these factors and labeled 30 decontextualized statements for sentiment, including meta-data on labeling rationale and uncertainty.

Our results show: (1) Sentiment perception is only moderately stable within individuals, and label changes concentrate on ambiguity-prone statements; (2) Correlation-level signals are small and do not survive global multiple-testing correction; (3) In statement-level repeated-measures models (GEE), higher mood trait and reactivity are associated with more positive (and less neutral) labeling, while predictors of negative labeling are weaker and at most trend-level (e.g., task conflict); (4) We find no clear evidence of systematic project-phase effects. Overall, sentiment perception varies within persons and is strongly statement-dependent.

Although our study was conducted in an academic setting, the observed variability and ambiguity effects suggest caution when interpreting sentiment analysis outputs and motivate future work with contextualized, in-project communication.}


\jnlcitation{\cname{%
\author{Obaidi M.},
\author{Herrmann M.},
\author{Martensen J.},
\author{Klünder J.}, and
\author{Schneider K.}}.
\ctitle{Exploring Indicators of Developers’ Sentiment Perceptions in Student Software Projects.}
\cjournal{\it Softw Pract Exp.}
\cvol{2024;00(00):1--18}.}

\maketitle



\section{Introduction}
The success rate of software projects can be increased by a good mood among developers~\cite{graziotin2014happy}. Developers in a positive mood are more productive and outperform those in a negative mood, particularly in problem-solving tasks~\cite{graziotin2014happy,graziotin2015you,10.1002/smr.1673}. Accordingly, negative emotions within a team, if ignored or neglected, pose a risk to project success. A developer's emotional state can be influenced by many factors, including life circumstances, personality, stress, and health~\cite{parrott2001emotions,shaver1987emotion,russell1980affect}. 

One crucial factor related to mood is communication. Developers may perceive the same statement very differently~\cite{herrmannSentiSurvey22}, which can complicate collaboration. For instance, a message intended as positive might be interpreted negatively due to missed cues like irony, sarcasm, or ambiguous language~\cite{herrmannSentiSurvey22}. Conversely, a developer's current mood can affect the tone of the messages they write~\cite{schroth2022potential,specht2024sentiment}, possibly triggering emotional contagion and shaping the emotional climate of the entire team. Communication, therefore, plays a central role in shaping how developers feel and interact—and must be handled with care.

However, perception of communication is influenced by many contextual and personal factors. Herrmann et al.~\cite{herrmannSentiSurvey22} showed that developers differ widely in how they interpret emotionally charged statements, possibly due to mood, experience, or individual differences.

In this paper, we investigate which factors shape the perception of sentiment in software-related communication. Specifically, we study 81 computer science students participating in real software projects as part of a university course in which 204 students worked in 28 team-based projects. Students labeled communication statements as positive, neutral, or negative and self-reported mood, life satisfaction, and group dynamics at four project phases.

While our focus is on student developers, prior work suggests that studies in educational contexts can provide indications that may be relevant to professional software development, but transferability is not guaranteed and differences should be expected. For example, Pankaj et al.~\cite{pankaj2019softwareuniindustry} report similarities between students and professionals in several technical tasks, and Salman et al.~\cite{salman2015studentindustry} found comparable performance when both groups adopted unfamiliar development techniques such as test-driven development. Counsell et al.~\cite{Counsell2008studentindustry} similarly observed no significant differences in cohesion assessments in their setting, while also noting differences in how students and industry participants arrived at their judgments. At the same time, important differences remain: Clark et al.~\cite{clark2003studentindustry} observed that professional developers are often more extraverted and communicative than students, reflecting differences in motivation, experience, and team structures. Accordingly, we position our results primarily within the context of software engineering education and discuss potential implications for professional practice cautiously.

\textbf{Our study provides the following insights:} 
(1) Student developers' perceptions of the same statements change over time.
(2) Mood-related factors show \emph{small} associations: correlation-level signals are weak and do not survive global multiple-testing correction, while statement-level repeated-measures models suggest a shift mainly toward \emph{more positive (and less neutral)} labeling.
(3) Task conflict is weakly associated with more negative sentiment assignments (exploratory trend).
(4) Sentiment perception varies across individuals and statements, and is sensitive to context and time.

These findings suggest that sentiment perception is dynamic—even within the same individual. Therefore, researchers and practitioners should account for this variability to better understand team communication and foster collaboration.

\textit{Outline.} The rest of the paper is structured as follows: In Section~\ref{sec:background}, we present related work and background details. The study design is introduced in Section~\ref{sec:research}. Section~\ref{sec:results} summarizes the results, which are discussed in Section~\ref{sec:discussion}, before concluding in Section~\ref{sec:conclusion}.

\section{Background and Related Work}
\label{sec:background}

In this section, we present related work on emotion models and sentiment analysis in software engineering.

\subsection{Emotion Frameworks}
Emotion classification frameworks categorize higher-level emotions, with lower-level emotions assigned to these categories~\cite{shaver1987emotion}. Shaver et al.~\cite{shaver1987emotion} developed one such framework based on data compiled by psychology students.

Russell's \textit{Core Affect} model describes mood as a “core affect with no object”~\cite{russell2003CoreAA}. Core affect results from the mixture of two dimensions: pleasure or displeasure and sleepy or active~\cite{russell2003CoreAA}. The specific feeling arising from this combination can be determined using a circumplex model~\cite{russell1980affect}, which Russell also developed for core affect~\cite{russell1980affect}. In this model, two dimensions combine to generate a specific emotional experience.

Another model developed by Russell in collaboration with Mehrabian~\cite{mehrabian1974approach} introduces a third dimension, \enquote{dominance}, in addition to the pleasure and arousal dimensions from the previous model. Similarities between models often appear in the selection of dimensions, such as \enquote{pleasure} and \enquote{arousal}~\cite{mehrabian1974approach,osgood1957measurement,watson1985toward}. Although these dimensions are sometimes labeled differently, they share consistent content across models. Examples include the models by Osgood et al.~\cite{osgood1957measurement} and Watson and Tellegen~\cite{watson1985toward}, both of which use these two dimensions.

To measure mood, mood scales are used, such as the one developed by Bohner et al.~\cite{bohner1991stimmungs}, a German adaptation of the \enquote{Mood Survey} by Underwood and Froming~\cite{underwood1980mood}. This scale measures general mood, referred to as an “enduring mood”~\cite{bohner1991stimmungs},  and can be divided into two subscales: one for capturing enduring mood and one for assessing a person's emotional reactivity~\cite{bohner1991stimmungs}. Reactivity measures the intensity of mood changes. The scale consists of 15 statements, rated on a 7-point scale~\cite{bohner1991stimmungs,bohner1997stimmungsskala}, with negative polarity items reversed during evaluation to ensure accurate results~\cite{bohner1991stimmungs,bohner1997stimmungsskala}. After reversal, higher values represent a better mood or stronger emotional reactivity~\cite{bohner1991stimmungs,bohner1997stimmungsskala}.

For assessing short-term affect, the Positive and Negative Affect Schedule (PANAS)~\cite{watson1988development} can be used. 
PANAS comprises 20 adjectives, with 10 items each for positive and negative affect, and is typically rated on a 5-point intensity scale using a specified time frame such as the past week~\cite{watson1988development}. 
German adaptations exist, for example by Krohne et al.~\cite{krohne1996untersuchungen} and as documented for the GESIS Panel by Breyer and Bl{\"u}mke~\cite{breyer2016panas-de}.

Beyond software engineering, research in psychology has long investigated how affective state can shape information processing and interpretation, including mood-congruent effects in judgment and memory~\cite{bower1981MoodMemory}.

\subsection{Sentiment Analysis in Software Engineering}

Recent systematic mapping studies and literature reviews have focused on sentiment analysis in software engineering, providing an overview of the datasets used in this field~\cite{obaidi2021sentimenttools,obaidi2022mappingstudy,linSentiSLR22}. One key finding is the subjectivity of sentiment annotation, which often leads to low agreement between annotators~\cite{obaidi2022mappingstudy,obaidi2022sentimenttoolsdata}.

Herrmann et al.~\cite{herrmannSentiSurvey22,herrmann2022subjectivitydata} explored the subjectivity of sentiment perception by analyzing statements from Stack Overflow and GitHub, using two established datasets~\cite{lin18sentiment,novielligold.2020}. In their study, computer scientists annotated 100 statements based on their own perception. The results revealed that 62.5~\% of participants' annotations aligned with the scientific authors' annotations~\cite{herrmannSentiSurvey22}.

Herrmann et al.~\cite{herrmann2025different-perceptions} investigated software developers' perceptions of sentiment in project communication by analyzing how development teams interpret these sentiments. Using hierarchical cluster analysis on 94 developers, they identified two distinct groups based on sentiment ratings. Interestingly, about 65\% of statements showed significant perception differences between the groups, with demographic factors such as age and experience not accounting for these differences.

Schroth et al.~\cite{schroth2022potential} introduced a concept for real-time sentiment analysis in software projects, evaluating it through an experiment with industry participants. They investigated how such a system could be integrated into developers' daily work to monitor and analyze sentiments in real-time.

Herrmann et al.~\cite{sentianalyzerreport2022,herrmann-kluender-2021} developed a voting classifier called SEnti-Analyzer, which combines three machine learning methods to analyze German texts, especially in the context of meetings~\cite{herrmann-kluender-2021}. This classifier integrates multiple sentiment analysis tools to improve accuracy.

Obaidi et al.~\cite{obaidi22cross} combined several pre-trained tools with different datasets to address the issue of poor cross-platform performance in sentiment analysis. They found that while this approach offers robust overall performance, individual tools still tend to perform slightly better on specific datasets. One of their suggestions is to conduct a more detailed analysis of the datasets used for training.

Novielli et al.~\cite{novielligold.2020} developed a gold standard dataset based on GitHub discussions, consisting of over 7,000 statements. These statements were annotated for emotions using Shaver's emotion model~\cite{shaver1987emotion}, and polarity labels were assigned based on the identified emotions.

Lin et al.~\cite{lin18sentiment} collected a dataset of 1,500 discussions related to Java programming on Stack Overflow. Five authors collaboratively labeled the dataset using a custom-built web application. However, their paper did not reference any specific emotion model or labeling guidelines.

Herrmann et al.~\cite{herrmann2025montecarlo} address the challenge of sentiment interpretation variability within software teams and propose a mathematical model to assess how representative a team's sentiment perception is when not all members participate in a mood survey. Through a Monte Carlo experiment with 45 developers, they demonstrate that omitting even one member from an average-sized team can significantly distort the perceived team mood, highlighting the importance of capturing diverse sentiment perceptions to avoid misjudging team atmosphere.

Obaidi et al.~\cite{obaidi2025germandataset} address the limited availability of non-English gold-standard resources for sentiment analysis in software engineering by providing a German dataset of 5,949 developer statements from a German developer forum. The dataset is annotated with six basic emotions following Shaver's model~\cite{shaver1987emotion} and shows high agreement and reliability among German-speaking raters. Their evaluation of existing German sentiment analysis tools indicates insufficient performance for software engineering texts, motivating dedicated domain-specific resources and models.

Obaidi et al.~\cite{obaidi2025trustworthy} study the trustworthiness and context dependence of sentiment analysis tools across platforms by analyzing linguistic and statistical characteristics of 10 developer communication datasets and evaluating 14 tools. They report substantial cross-platform performance variation and propose a feature-based mapping approach (including a questionnaire) to recommend suitable tools for new datasets without requiring labeled data. Their results highlight that even strong models (e.g., transformer-based approaches) are not universally optimal and that dataset characteristics are critical for reliable sentiment analytics in software engineering.

\subsection{Research Gap and Contribution}
Prior studies on sentiment analysis in software engineering have highlighted the complexity and subjectivity of emotion perception in developer communication~\cite{herrmannSentiSurvey22,lin18sentiment,novielligold.2020}. While these works demonstrate disagreement among annotators and the challenges of consistent labeling, most studies have focused on static, cross-sectional data, often relying on third-party annotations.

A key limitation in prior work is the lack of longitudinal analysis—little is known about how the same developer's sentiment perception changes over time. Furthermore, most research has not systematically investigated the role of psychological and contextual factors, such as mood or team conflict, in shaping sentiment interpretation.

Our study addresses these gaps by (1) conducting a longitudinal survey over four time points, (2) integrating both emotional and project-related factors (e.g., PANAS, group conflicts, life satisfaction), and (3) analyzing intra-individual sentiment perception dynamics. In contrast to prior research, we explore how situational and individual factors may jointly influence how developers interpret written communication over time.

\section{Study Design}
\label{sec:research}
In this section, we present our study design, including the research objective, research questions, survey instrument, and data collection and analysis procedures.

\subsection{Research Objective and Research Questions}
Our overall objective is to \textit{analyze which factors are associated with whether developers perceive a textual statement as positive, negative, or neutral}. This undertaking contributes to (1) shedding light on the variety of factors influencing mood, (2) raising awareness of both subjective and objective factors that affect communication perception, and (3) highlighting the diversity of psychological aspects within development teams. 

To achieve this objective, we pose the following research questions:

\vspace{6pt}\noindent\textbf{RQ1:} \textit{How stable is the perception of individual developers towards textual statements?}
\newline\noindent
The first step in achieving our research goal is to examine whether a developer’s perception remains consistent when evaluating the same statements repeatedly over a predefined period. Answering this question provides a foundation for the subsequent research questions. If perception changes over time (intra-personal variation), we can analyze these changes in more detail. Otherwise, we are limited to analyzing differences between developers (inter-personal variation). 

\vspace{6pt}\noindent\textbf{RQ2:} \textit{How are team-internal (psychological) aspects, such as mood or conflicts, related to how developers perceive text-based communication?}
\newline\noindent
This research question explores various psychological factors that may influence developers' perceptions of text-based statements. These factors are subjective by nature, and analyzing how they vary among developers highlights the diversity of perceptions due to subjectivity.

\vspace{6pt}\noindent\textbf{RQ3:} \textit{How are external factors, such as project phases, related to how developers perceive text-based communication?}
\newline\noindent
In line with RQ2, we also aim to analyze objective factors, such as project phases (e.g., the time before a deadline or near project completion). This question addresses how these objective factors, which may affect the entire team, influence perceptions. By understanding these influences, teams can be more mindful of maintaining polite and friendly communication during critical phases.

\subsection{Survey Instrument}
We used the survey methodology of Robson and McCartan~\cite{robson2016real} for data collection, implementing the survey as an online questionnaire using \textit{LimeSurvey}. 
We selected a survey-based approach because our research focuses on subjective, individual-level perceptions that are best captured through self-report instruments. This is consistent with prior research on sentiment perception in software engineering, such as Herrmann et al.~\cite{herrmannSentiSurvey22}, who successfully used surveys to examine how developers interpret the sentiment of textual statements. Surveys are also well-suited for collecting data across different project phases and for assessing personal factors such as mood, life satisfaction, and perceived group dynamics—variables that are difficult to capture through observational or experimental methods.

\subsubsection{Survey Structure}
The survey consisted of seven groups of questions, totaling 13 questions. The groups and their number of questions are summarized in Table~\ref{tab:survey}. An overview of the response formats and scale anchors used for the key constructs (mood traits and states, life circumstances, and group dynamics) is provided in Table~\ref{tab:scales}.

\begin{table}[htbp]
\caption{Survey Structure}
\label{tab:survey}
\begin{tabularx}{\columnwidth}{lXc}
\toprule
& \textbf{Category} & \textbf{Questions} \\
\midrule
\textbf{(1)} & Demographics & 5 \\
\textbf{(2)} & \textit{Long-term} mood trait analysis & 1 \\
\textbf{(3)} & \textit{Short-term} mood state analysis & 1 \\
\textbf{(4)} & Life circumstances & 1 \\
\textbf{(5)} & Sentiment perception of statements & 1 \\
\textbf{(6)} & Annotation criteria (labeling rationale) & 3 \\
\textbf{(7)} & Group dynamics within the software project & 1 \\
\bottomrule
\end{tabularx}
\end{table}

\begin{table}[htbp]
\caption{Overview of response scales}
\label{tab:scales}
\begin{tabularx}{\columnwidth}{lX}
\toprule
\textbf{Construct} & \textbf{Scale anchors}\\
\midrule
Life circumstances~\cite{SurveyLeben2023,Bundesregierung2020,HurrelmannRichter2022}
  & 1 = totally dissatisfied, 5 = very satisfied \\
Mood trait~\cite{underwood1980mood,bohner1991stimmungs,bohner1997stimmungsskala}
  & 1 = does not apply at all, 7 = applies completely \\
PANAS~\cite{watson1988development,krohne1996untersuchungen}
  & 1 = not at all, 5 = extremely (past week) \\
Group dynamics (conflicts)~\cite{jehn1995multimethod,lehmann2011intragroup}
  & 1 = never, 6 = very often \\
\bottomrule
\end{tabularx}
\end{table}

\subsection{Survey Items}
\textit{Demographics} included questions about participants' familiarity with English and details regarding their project experience.

For the \textit{mood trait analysis}, we used the German version of the mood survey by Underwood and Froming~\cite{underwood1980mood}, provided by Bohner et al.~\cite{bohner1991stimmungs}. Items were rated using the response scale shown in Table~\ref{tab:scales}. This scale assesses typical, long-term mood~\cite{underwood1980mood} by asking participants to respond to 15 items, as listed in Table~\ref{tab:mood-trait}. According to Bohner et al.~\cite{bohner1991stimmungs}, items with negative polarity must be reversed for accurate evaluation. Higher scores on this scale indicate a generally positive mood or a highly reactive person who frequently experiences mood changes. The maximum possible score is 63 for the mood subscale and 42 for reactivity.

\begin{table}[htbp]
\caption{Mood Survey Items by Underwood and Froming~\cite{underwood1980mood}}
\label{tab:mood-trait}
\begin{tabularx}{\columnwidth}{X}
\toprule
\textbf{Items for Mood} \\
\midrule
\textbullet~I usually feel quite cheerful. \\
\textbullet~I'm frequently ``down in the dumps''.\\
\textbullet~I generally look at the sunny side of life.\\
\textbullet~I'm not often really elated. \\
\textbullet~I usually feel as though I'm bubbling over with joy. \\
\textbullet~I consider myself a happy person.\\
\textbullet~Compared to my friends, I think less positively about life in general. \\
\textbullet~I am not as cheerful as most people.\\
\textbullet~My friends often seem to feel I am unhappy. \\
\toprule
\textbf{Items for Reactivity}\\
\midrule
\textbullet~I may change from happy to sad and back again several times in a single week. \\
\textbullet~Compared to my friends, I'm less up and down in my mood states.\\
\textbullet~Sometimes my moods swing back and forth very rapidly. \\
\textbullet~My moods are quite consistent. They almost never vary. \\
\textbullet~I'm a very changeable person. \\
\textbullet~I'm not as ``moody'' as most people I know.\\
\bottomrule
\end{tabularx}
\end{table}

For the \textit{mood state analysis}, we used the Positive and Negative Affect Schedule (PANAS)~\cite{watson1988development} in the German version by Krohne et al.~\cite{krohne1996untersuchungen}. 
This instrument measures positive and negative affect with 10 items each, rated on a 5-point ordinal intensity scale (1 = not at all, 5 = extremely) referring to the past week. 
Table~\ref{tab:panas} summarizes the items for positive and negative affect, asking participants to indicate to what extent they experienced these feelings during the past week.

\begin{table}[!htbp]
	\caption{Items related to positive and negative affect~\cite{watson1988development}}
	\label{tab:panas}
	\begin{tabularx}{\columnwidth}{XXXX}
		\toprule 
		\multicolumn{2}{l}{\textbf{Positive Affect}} & \multicolumn{2}{l}{\textbf{Negative Affect}}\\
		\midrule
			\textbullet~active & \textbullet~determined & \textbullet~afraid &\textbullet~irritable \\
			\textbullet~attentive & \textbullet~inspired & \textbullet~ashamed &\textbullet~jittery \\
			\textbullet~alert & \textbullet~interested & \textbullet~distressed &\textbullet~nervous \\
			\textbullet~excited & \textbullet~proud & \textbullet~guilty &\textbullet~scared \\
			\textbullet~enthusiastic & \textbullet~strong & \textbullet~hostile &\textbullet~upset\\
		\bottomrule 
	\end{tabularx} 
\end{table}

For \textit{life circumstances}, we asked the participants to rate their satisfaction with nine aspects of their daily lives on a 5-point ordinal scale, as summarized in Table~\ref{tab:scales}. Table~\ref{tab:lifec} lists these aspects, which were drawn from studies on quality of life in Germany conducted by three institutions~\cite{SurveyLeben2023,Bundesregierung2020,HurrelmannRichter2022}. They emerged from a survey of the Federal Statistical Office on quality of life~\cite{SurveyLeben2023}, a report of the Federal Government on quality of life~\cite{Bundesregierung2020}, and a contribution to health of the Federal Center for Health Education in Germany~\cite{HurrelmannRichter2022}. Participants could additionally select \enquote{Not applicable} for individual items. We treated these responses as missing and computed $liv$ as the mean across answered items (range 1--5).

To assess reliability, we computed Cronbach's $\alpha$ in two ways. Listwise deletion on complete cases yielded $\alpha = 0.813$ ($n = 101$). A pairwise approach based on pairwise covariances yielded $\alpha = 0.788$. Both indicate acceptable internal consistency~\cite{kotz2005encyclopedia}.

\begin{table}[!htbp]
	\caption{Items related to the life circumstances~\cite{SurveyLeben2023,Bundesregierung2020,HurrelmannRichter2022}}
	\label{tab:lifec}
	\begin{tabularx}{\columnwidth}{lX}
		\toprule 
		\textbf{Circumstance} & \textbf{Examples}\\
		\midrule
			Work & Relationship with colleagues, enjoyment of work\\
			Life situation & Feel-good factor, sufficient privacy \\
			Finance & Financial independence, monthly income compared to expenditure \\
			Self-determination & Scope for own decisions, proportion of own decisions\\
			Safety & Feeling safe when leaving the house, social security \\
			Study & Progress in studies, interest in academic content \\
			Health & Fitness, physical well-being \\
			Social environment & Friends, family, other social contacts \\
 			Self-realization & Enjoyment of own activities, achieving own goals \\
		\bottomrule 
	\end{tabularx} 
\end{table}

To analyze the \textit{sentiment perception of text-based statements}, we asked the participants to label a total of 30 statements as positive, negative, or neutral.
These statements were taken from GitHub and Stack Overflow datasets, with half sourced from Novielli et al.~\cite{novielli2020githubgold} and the other half from Lin et al.~\cite{lin18sentiment}. Examples of these statements are in Table~\ref{tab:example_statements}. Each dataset provided baseline labels, which were used for comparison with participants' perceptions. We randomly selected 30 statements (15 from GitHub, 15 from Stack Overflow) and ensured an even distribution of positive, neutral, and negative labels (5 of each from both datasets). We randomized the question order to mitigate potential biases.

\begin{table}[htbp]
\caption{Example Statements with Ground Truth Labels}
\label{tab:example_statements}
\begin{tabularx}{\columnwidth}{lXl}
\toprule
\textbf{Source} & \textbf{Statement} & \textbf{Label} \\
\midrule
GitHub & I know a lot of server admins will be happy about that. & Positive \\
GitHub & How would you prefer it to be implemented? & Neutral \\
GitHub & quit spamming my notifications please, kthkbye & Negative \\
Stack Overflow & And it works like a charm now SMILE\_FACE. & Positive \\
Stack Overflow & which someone converted to a JSON string as follows. & Neutral \\
Stack Overflow & Don't use. & Negative \\
\bottomrule
\end{tabularx}
\end{table}

For the \textit{annotation criteria}, we asked participants to specify which factors influenced their label selection, such as content, tone, or the use of emoticons. They were also given a free-text field to elaborate on their choices. Participants then rated their confidence in their label selection on a 5-point ordinal scale. To further investigate uncertainties, we asked if they felt there were too many possible interpretations (e.g., content, tone, and emoticons pointing to different labels), detected contradictory emotions, or found it difficult to define one label. Again, a text field was provided for additional explanations.

Finally, we gathered information on group dynamics using Jehn's intragroup conflict scale~\cite{jehn1995multimethod}, in its German adaptation~\cite{lehmann2011intragroup}. Participants responded using the 6-point frequency scale shown in Table~\ref{tab:scales}. The items are presented in Table~\ref{tab:conflicts}. 

\begin{table}[!htbp]
	\caption{Items related to relationship and task-related conflicts~\cite{jehn1995multimethod}}
	\label{tab:conflicts}
	\begin{tabularx}{\columnwidth}{XX}
		\toprule 
		\textbf{Relationship conflicts} & \textbf{Task-related conflicts}\\
		\midrule
		\textbullet~How much friction is there among members in your work unit? & \textbullet~How often do people in your work unit disagree about opinions regarding the work being done?\\
		\textbullet~How much are personality conflicts evident in your work unit? & \textbullet~How frequently are there conflicts about ideas in your work unit?\\
		\textbullet~How much tension is there among members in your work unit? & \textbullet~How much conflict about the work you do is there in your work unit?  \\
		\textbullet~How much emotional conflict is there among members in your work unit? & \textbullet~To what extent are there differences of opinion in your work unit? \\
		\bottomrule 
	\end{tabularx} 
\end{table}

All the instruments and scales used in this study were chosen due to their publication in well-regarded, peer-reviewed research and their widespread acceptance in related fields. Their established reliability and validity make them well-suited for investigating the psychological and contextual factors influencing sentiment perception in software projects. By leveraging these well-documented tools, we aimed to ensure methodological rigor and replicability in our analysis.

In addition to assigning sentiment labels, participants reported meta-information about their labeling decisions at the questionnaire level (i.e., for the full set of 30 statements).
Specifically, they indicated which criteria influenced their choices (content, tone, emoticons, and an optional free-text field), and they reported their confidence in the assigned labels on a 5-point scale.
To further characterize uncertainty, participants selected whether labeling was difficult due to (i) too many possible interpretations, (ii) contradictory emotions, or (iii) no label fitting well, and could add an optional free-text explanation.

We analyzed these meta-data descriptively by reporting frequencies of selected criteria and uncertainty reasons.
To relate self-reported confidence to labeling behavior, we computed two additional questionnaire-level measures: (a) \emph{mismatch rate}, the proportion of statements where a participant's label differed from the dataset's reference label, and (b) \emph{label entropy}, the entropy of the participant's own label distribution over \{positive, neutral, negative\} within a questionnaire.
We tested associations between confidence and these measures using Spearman correlation.
Finally, we compared confidence distributions between participants who did vs.\ did not report specific uncertainty reasons using Mann--Whitney U tests and applied Holm and Benjamini--Hochberg corrections across these comparisons.

\subsection{Data Collection}
\begin{figure*}[t]
\includegraphics[width=\textwidth]{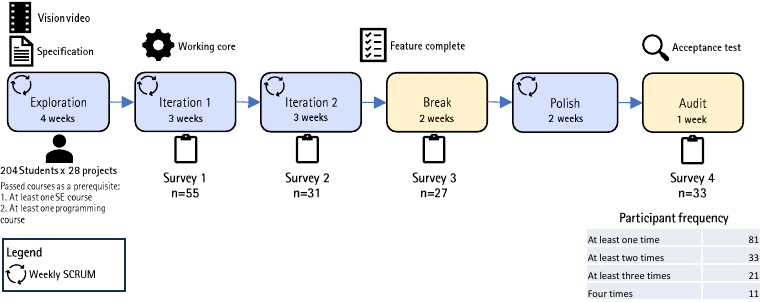}
\caption{Overview of the ``Software Project'' process, conducted from October 2021 to January 2022}
\label{fig:swp}
\end{figure*}

We conducted the survey during the winter semester 2021/2022 in a course called ``Software Project'' at Leibniz University Hannover in Germany. 
Figure~\ref{fig:swp} provides an overview of the ``Software Project'' process. This course is part of the bachelor's program for computer science students, where teams of six to nine students develop software over a period of 15 weeks for real customers. Students participating in this project must have already passed core modules in software engineering and programming, ensuring that each participant has prior programming experience and is familiar with software engineering principles. The project includes several phases, as visualized in Figure~\ref{fig:swp}: exploration, two iterations, a holiday break, polish, and final audit. Each phase requires students to deliver specific outputs, such as a project specification, a vision video, or software increments.

Although this setting differs from professional software development, it closely resembles real-world conditions through the use of external customers, iterative development, and structured project management. As such, it provides a relevant educational environment to study developer behavior in a controlled, yet authentic, team-based context.

In 2021/2022, 204 students participated in the course, working in teams across 28 software projects, all of which passed the final audit successfully. We invited students to participate in the survey via email and through their student tutors. Participation was voluntary.  

Prior work suggests that structured project courses can be useful settings to study developer behavior in controlled team contexts, while differences to professional environments should be expected~\cite{Counsell2008studentindustry,salman2015studentindustry}.

\begin{figure}[ht]
\centering
\includegraphics[scale=1.0]{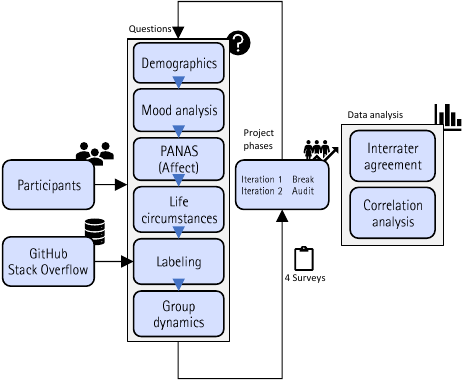}
\caption{Overview of our study design}
\label{fig:studydesign}
\end{figure}

The survey was administered four times in data-collection windows (survey rounds) aligned with key project milestones: after iteration~1, after iteration~2, during the holiday break, and during the final audit, as shown in Figure~\ref{fig:studydesign}. Each round was open for a time window, and teams progressed at slightly different paces. Therefore, the project phase recorded at the time a questionnaire was completed can differ from the nominal phase targeted by that round.

We received a total of 146 completed questionnaires across the four rounds: 55 in Round~1, 31 in Round~2, 27 in Round~3, and 33 in Round~4. For phase-based analyses, we use the phase value recorded per submission. Across all 146 submissions, phase counts were iter1 ($n=49$), iter2 ($n=31$), break ($n=33$), and audit ($n=33$).

In total, 81 students participated in the survey at least once. Of these, 33 submitted at least two questionnaires, 21 submitted at least three questionnaires, and 11 submitted four questionnaires. A small number of participants submitted the questionnaire twice within the same round. We retained these complete submissions and capture repeated participation via the participation index.

\begin{table*}[htb]
\caption{Overview of variables used for our analysis. $p$ stands for participant.}
\label{tab:variables}
\begin{tabularx}{\textwidth}{lp{4.2cm}lrX}
\toprule
\textbf{Variable} & \textbf{Name} & \textbf{Scale} & \textbf{Range} & \textbf{Description} \\
\midrule
$mo$ & Mood trait & ordinal & 9..63 & Sum of all mood items (Table~\ref{tab:mood-trait}) rated by $p$ \\
$rea$ & Reactivity trait & ordinal & 6..42 & Sum of all reactivity items (Table~\ref{tab:mood-trait}) rated by $p$ \\
$pos$ & Positive mood state & ordinal & 10..50 & Sum of PANAS positive affect items (Table~\ref{tab:panas}) rated by $p$ \\
$neg$ & Negative mood state & ordinal & 10..50 & Sum of PANAS negative affect items (Table~\ref{tab:panas}) rated by $p$ \\
$liv$ & Life circumstances & interval & 1..5 & Mean score across answered life circumstance items (Table~\ref{tab:lifec}). ``Not applicable'' treated as missing \\
$rel$ & Relationship conflicts & ordinal & 4..24 & Sum of relationship conflict items (Table~\ref{tab:conflicts}) rated by $p$ \\
$tas$ & Task-related conflicts & ordinal & 4..24 & Sum of task-related conflict items (Table~\ref{tab:conflicts}) rated by $p$ \\
\midrule
$num_{pos}$ & Number of positive labels & interval & 0..30 & Number of positive labels per questionnaire \\
$num_{neu}$ & Number of neutral labels & interval & 0..30 & Number of neutral labels per questionnaire \\
$num_{neg}$ & Number of negative labels & interval & 0..30 & Number of negative labels per questionnaire \\
$\kappa(p)$ & Intrarater agreement of participant $p$ & interval & [-1; 1] & Self-agreement across repeated questionnaires for $p$, calculated with Fleiss' $\kappa$ \\
\midrule
$phase$ & Project phase (categorical) & nominal & \makecell[t]{\{iter1, iter2,\\ break, audit\}} & Project phase at which a questionnaire was completed \\
$phase_{ord}$ & Project phase (temporal order) & ordinal & 1..4 & Ordinal coding of project phases in chronological order (used for correlation tests in RQ3) \\
$idx(p,r)$ & Participation index & ordinal & 1..4 & The $r$-th submitted questionnaire of participant $p$ (ordered by submission time) \\
$eng$ & English proficiency & ordinal & 1..5 & Self-rated English proficiency (1--5). Native speakers coded as 5 \\
\bottomrule
\end{tabularx}
\end{table*}

\begin{figure}[t]
  \centering
  \includegraphics[scale=1.0]{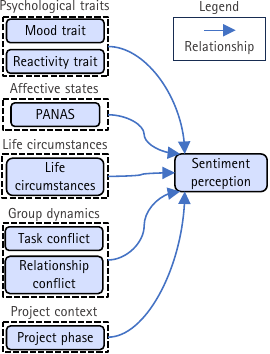}
  \caption{Conceptual path diagram of the examined associations between individual and contextual factors (psychological traits and states, life circumstances, group dynamics, and project context) and sentiment perception of text-based statements}
  \label{fig:pathdiagram}
\end{figure}

Figure~\ref{fig:pathdiagram} summarizes the conceptual structure of our study and the examined associations between individual and contextual factors and sentiment perception in text-based statements. The diagram is intended as an orienting overview (non-causal).

\subsection{Data Analysis}
\label{sec:dataanalysis}

\subsubsection{Preprocessing and sample handling}
The survey data was collected using LimeSurvey\footnote{\url{https://www.limesurvey.org/de}} and then pre-processed for analysis. Incomplete surveys were excluded from the analysis. In a small number of cases, a participant submitted the questionnaire more than once within the same phase. We retained these complete submissions and treat them as separate measurement occasions (captured by the participation index).

Table~\ref{tab:variables} outlines the variables used in our analysis. To answer our research questions, we examined the relationships between these variables, focusing particularly on how they relate to the perception of text-based statements.

\subsubsection{Within-participant stability}
To quantify within-participant self-agreement on label assignments across repeated questionnaires, we computed Fleiss' $\kappa$~\cite{fleiss1971measuring} for participants who completed at least two questionnaires. Conceptually, we treat each questionnaire submission as one rating occasion for the same fixed set of statements within a participant and use $\kappa$ as a chance-corrected agreement measure for multiple rating occasions with categorical labels.

\subsubsection{Bivariate hypothesis testing (Spearman correlations)}
Since most data were collected on ordinal scales, we used Spearman's rank correlation coefficient ($\varrho$)~\cite{spearman1904rho} to measure the strength and direction of the relationships between variables. For interpreting the correlation coefficients, we followed Cohen’s scale~\cite{Cohen1988}. We assessed statistical significance using a significance level of $\alpha = 0.05$ and report uncorrected $p$-values. Because some variables are missing for subsets of questionnaires (e.g., mood trait and reactivity), the effective sample size can vary by test. We therefore report the used $n$ per correlation (pairwise deletion).

We tested the hypotheses presented in Table~\ref{tab:merged_hypotheses}.

\begin{table*}[!htbp]
\caption{Null hypotheses and variables for RQ1--RQ3}
\label{tab:merged_hypotheses}
\begin{tabularx}{\textwidth}{lXp{4.3cm}}
\toprule
\textbf{Hypothesis} & \textbf{Description (There is no relation between \ldots)} & \textbf{Variables} \\
\midrule
H$1_0$ & \ldots intrarater agreement and psychological factors. & $\kappa(p)$, $mo$, $rea$, $pos$, $neg$, $liv$ \\
H$1.1_0$--H$1.5_0$ & This applies to mood trait, reactivity, positive affect, negative affect, or life circumstances, respectively. & \\ \midrule
H$2_0$ & \ldots the perceived polarity of statements and mood. & $num_{pos}$, $num_{neu}$, $num_{neg}$, $mo$ \\
H$2.1_0$--H$2.3_0$ & This applies to positive, neutral, or negative labels. & \\ \midrule
H$3_0$ & \ldots the perceived polarity of statements and reactivity. & $num_{pos}$, $num_{neu}$, $num_{neg}$, $rea$ \\
H$3.1_0$--H$3.3_0$ & This applies to positive, neutral, or negative labels. & \\ \midrule
H$4_0$ & \ldots the perceived polarity of statements and positive mood state. & $num_{pos}$, $num_{neu}$, $num_{neg}$, $pos$ \\
H$4.1_0$--H$4.3_0$ & This applies to positive, neutral, or negative labels. & \\ \midrule
H$5_0$ & \ldots the perceived polarity of statements and negative mood state. & $num_{pos}$, $num_{neu}$, $num_{neg}$, $neg$ \\
H$5.1_0$--H$5.3_0$ & This applies to positive, neutral, or negative labels. & \\ \midrule
H$6_0$ & \ldots the perceived polarity of statements and life circumstances. & $num_{pos}$, $num_{neu}$, $num_{neg}$, $liv$ \\
H$6.1_0$--H$6.3_0$ & This applies to positive, neutral, or negative labels. & \\ \midrule
H$7_0$ & \ldots the perceived polarity of statements and relationship conflicts. & $num_{pos}$, $num_{neu}$, $num_{neg}$, $rel$ \\
H$7.1_0$--H$7.3_0$ & This applies to positive, neutral, or negative labels. & \\ \midrule
H$8_0$ & \ldots the perceived polarity of statements and task-related conflicts. & $num_{pos}$, $num_{neu}$, $num_{neg}$, $tas$ \\
H$8.1_0$--H$8.3_0$ & This applies to positive, neutral, or negative labels. & \\ \midrule
H$9_0$ & \ldots the perceived polarity of statements and project phase. & $num_{pos}$, $num_{neu}$, $num_{neg}$, $phase_{ord}$ \\
H$9.1_0$--H$9.3_0$ & This applies to positive, neutral, or negative labels across phases. & \\
\bottomrule
\end{tabularx}
\end{table*}

First, for RQ1, we examined whether within-participant self-agreement in sentiment labeling ($\kappa(p)$) is associated with psychological factors (H$1_0$).
Second, for RQ2, we tested whether the perceived polarity of statements (counts of positive, neutral, and negative labels per questionnaire) is associated with mood trait, reactivity, positive/negative affect, life circumstances, and intragroup conflicts (H$2_0$--H$8_0$).
Finally, for RQ3, we examined whether label counts vary across project phases (H$9_0$). For this purpose, we encoded project phase as an ordinal variable reflecting temporal order ($phase_{ord}$) for correlation tests, while treating $phase$ as categorical in multivariate models.
Table~\ref{tab:corr_global} shows the correlations that reached the conventional uncorrected threshold of $p<0.05$.

\subsubsection{Multiple-testing correction}
Because we computed a battery of Spearman correlations across the hypothesis set, we additionally corrected for multiple comparisons across \emph{all} correlation tests.
Specifically, we report (i) Holm correction controlling the family-wise error rate (FWER) and (ii) Benjamini--Hochberg correction controlling the false discovery rate (FDR).
We treat correlations that reach the conventional uncorrected threshold ($p<0.05$) but do not remain below the adjusted thresholds as exploratory trends.

\subsubsection{Attrition and dropout-bias analysis}
Because participation was voluntary, the dataset contains missing waves and an unbalanced panel (participants contributed between one and four questionnaires).
To assess whether dropout could bias longitudinal interpretations, we performed an attrition analysis on the participant level.
We defined each participant's \emph{baseline} as their first submitted questionnaire (smallest participation index. Ties resolved by submission timestamp) and quantified \emph{retention} as the number of submitted questionnaires per participant ($completed\_questionnaires \in \{1,2,3,4\}$).
Our primary comparison contrasted participants with exactly one submitted questionnaire ($=1$) against those with at least two submitted questionnaires ($\geq 2$).

We first tested whether the availability of mood-trait data (with\_mood: yes/no) was associated with retention using Fisher's exact test.
Second, we assessed whether baseline phase is associated with retention, acknowledging that later baselines structurally reduce the opportunity to participate in subsequent waves. Again, we used Fisher's exact test.
For baseline group differences in psychological and project-related measures (mood trait, reactivity, PANAS positive/negative affect, life circumstances, relationship and task conflicts, and English proficiency), we used Mann--Whitney U tests because most measures are ordinal and subgroup sizes are small.
To control the family-wise error rate within this attrition analysis, we applied Holm correction across all baseline comparisons.
Finally, we quantified the association between baseline negativity in labeling and retention using logistic regression with $ge2$ (submitted at least two questionnaires) as outcome.
As sensitivity analysis, we repeated the baseline comparisons restricting baselines to Iteration~1 to reduce structural opportunity effects.

As an additional robustness check, we fitted a participant-level retention model and inspected stabilized inverse-probability weights (IPW). Because the weights were moderate and predictors showed no clear association with retention, we did not apply IPW in the main analyses.

\subsubsection{Statement-level multivariate modeling (GEE)}
To complement the bivariate correlation analyses and to account for the repeated-measures and nested structure of our data, we additionally modeled sentiment perception at the statement level using generalized estimating equations (GEE).
Each observation corresponds to one labeling decision for one statement (positive/neutral/negative).
We fit the models on the subset of questionnaires for which mood-trait data were available (with\_mood=yes), resulting in $128 \times 30 = 3840$ statement labels from 68 distinct participants.
We restricted the analysis to this subset because mood trait and reactivity are central predictors in our research questions. Including questionnaires without these measures would require additional missing-data handling (e.g., imputation), which is beyond the scope of this study.

Because sentiment labels are imbalanced (neutral labels dominate), we estimated three separate one-vs-rest logistic GEE models with a logit link.
(i) \emph{positive} vs.\ not positive.
(ii) \emph{negative} vs.\ not negative.
(iii) \emph{neutral} vs.\ not neutral.
We accounted for repeated measures by clustering on participants and using robust (sandwich) standard errors, yielding population-averaged effect estimates.

To control for systematic differences between statements, we included statement fixed effects (factor: statement\_id).
To account for project timing and repeated participation, we included phase fixed effects (factor: phase) and each participant's participation index (order of submissions by the same participant), capturing potential learning or fatigue effects beyond project phase.
We further included English proficiency as a control variable (self-rated proficiency on a 1--5 scale. Native speakers coded as 5).

Predictors of interest were mood trait, reactivity, PANAS positive/negative affect, life circumstances, and relationship and task conflicts.
All continuous predictors were z-standardized. Thus, odds ratios correspond to a one-standard-deviation increase in the predictor.

\subsubsection{Statement-level heterogeneity (disagreement and flipping)}
To quantify statement-level heterogeneity, we computed two complementary descriptive metrics for each statement.

\emph{Overall disagreement across submissions.}
For each statement $s$, we aggregated all sentiment labels assigned across all submitted questionnaires (i.e., across \emph{submissions}. $n=146$ labels per statement. Total 30 statements).
Note that these 146 labels do not stem from 146 independent annotators: they include repeated measurements from the same participants (and, in a few cases, multiple submissions within the same wave), such that participants with more submissions contribute more labels.
Accordingly, this metric captures a mixture of between-participant differences and within-participant changes, and we interpret it as an \emph{overall} disagreement measure across the observed submissions.

We computed (i) the \emph{majority share}, i.e., the proportion of labels belonging to the most frequent class, and (ii) Shannon entropy over the three labels $\{pos, neu, neg\}$:
$H(s) = -\sum_{c \in \{pos,neu,neg\}} p_{s,c}\log_2(p_{s,c})$,
where $p_{s,c}$ is the proportion of label $c$ for statement $s$.
Higher entropy indicates more mixed perceptions (maximum $\log_2(3)\approx 1.585$).

\emph{Within-participant instability (flip rate).}
To assess whether the same individual changes their perception of a given statement over time, we focused on participants who completed at least two questionnaires ($n=33$).
For each statement, we computed the \emph{flip rate}, defined as the proportion of these participants who assigned different labels to the same statement across their questionnaires (i.e., not constant over time).
We additionally computed a \emph{strong flip rate}, counting participants who switched between positive and negative at least once (pos$\leftrightarrow$neg), regardless of neutral assignments.
Finally, we examined the association between statement-level entropy (computed across submissions) and flip rate to assess whether statements with more mixed overall labeling also tend to be less stable within individuals.

\section{Results}
\label{sec:results}
In total, the survey was completed 146 times across four data-collection windows (survey rounds). The first round had 55 responses, the second had 31, the third had 27, and the final round had 33 responses. A total of 81 different participants took part in at least one round. The dataset is publicly available at \href{https://doi.org/10.5281/zenodo.18062116}{Zenodo}~\cite{obaidi2025swpDataset}.

Importantly, survey rounds and project phases are not identical in our data.
Each survey round was open for a time window, and teams progressed at slightly different paces. Therefore, the project phase recorded at the time a questionnaire was completed can differ from the nominal phase targeted by that round.
For analyses involving project phase, we therefore use the phase value recorded per submission.
Across all 146 submissions, phase counts were: iter1 ($n=49$), iter2 ($n=31$), break ($n=33$), and audit ($n=33$).
This distinction also explains why survey-round counts (e.g., Round~1: $n=55$, Round~3: $n=27$) do not match phase counts (e.g., iter1: $n=49$, break: $n=33$).

\subsection{Descriptive Analysis}
In the following, we present descriptive results of the variables used in our study. 

\subsubsection{Traits}
Mood and reactivity were measured using the mood survey by Underwood and Froming~\cite{underwood1980mood}. As summarized in Table~\ref{tab:variables}, the survey provides insights into the mood trait (variable $mo$) and the reactivity trait (variable $rea$).
To avoid over-counting repeated questionnaires, we report trait descriptives based on each participant's first available submission with mood-trait data (i.e., baseline per participant; $n=68$).
The mood subscale shows a median of 39 ($n = 68$, $min = 11$, $max = 59$). The median is above half of the possible points, indicating a tendency toward a more positive mood among the students. However, individual scores show substantial variability.
The reactivity subscale has a median of 19.5 ($n = 68$, $min = 10$, $max = 41$), just below half of the maximum possible points, again with notable variability.

\subsubsection{Mood State}
Mood state was assessed using the PANAS scale, which measures both positive and negative affect. Table~\ref{tab:panasdistr} summarizes the median PANAS scores for both positive and negative affect, along with the distribution of ranges, maxima, and minima across the four surveys.

\begin{table}[htbp]
\caption{Distribution of PANAS scores for positive and negative affect across surveys.}
\label{tab:panasdistr}
\begin{tabularx}{\columnwidth}{XXXXXXl}
\toprule
\multicolumn{2}{l}{\multirow{3}{*}{\textbf{Affect}}} & \multicolumn{5}{c}{\textbf{Survey}} \\
                            &         & \textbf{{\o}} & \textbf{1} & \textbf{2} & \textbf{3} & \textbf{4}\\ 
                            & & n=146 & n=55 & n=31 & n=27 & n=33\\
                            \midrule
\multirow{4}{*}{Positive}   & median  & 32          & 33       & 30       & 31.5     & 32       \\
                            & $\max$  & 46          & 45       & 43       & 44       & 46       \\
                            & range   & 31          & 26       & 28       & 21       & 28       \\
                            & $\min$  & 15          & 19       & 15       & 23       & 18       \\ \midrule
\multirow{4}{*}{Negative}   & median  & 22          & 23       & 23       & 21.5     & 21       \\
                            & $\max$  & 50          & 45       & 50       & 38       & 45       \\
                            & range   & 40          & 34       & 39       & 25       & 35       \\
                            & $\min$  & 10          & 11       & 11       & 13       & 10       \\ \bottomrule
\end{tabularx}
\end{table}

The median values for both positive and negative affect are relatively consistent across all four surveys. The variation of median scores across surveys is small (Positive: 30--33; Negative: 21--23). Positive affect is slightly above the scale midpoint (30), indicating a generally higher level of positive affect. In contrast, negative affect is well below the midpoint, suggesting comparatively low levels of negative affect overall.

Interestingly, the value range within surveys (difference between maximum and minimum) for negative affect is consistently wider than for positive affect, which is largely due to the minimum values for positive affect being higher (15) compared to negative affect (10). This suggests that while participants reported more stable positive affect, their negative affect experiences varied more significantly.

\subsubsection{Life Circumstances}
Life circumstances ($liv$) were operationalized as the mean satisfaction score across the answered life items (range 1--5), treating \enquote{Not applicable} as missing.
Across the four rounds, central tendencies were stable: the overall median was 3.56, and round-specific medians ranged from 3.44 to 3.56.
While the median remained stable, individual mean scores varied substantially (observed range 1.33--4.56).

For transparency regarding missingness, 101 responses contained valid values for all 9 items, 40 responses for 8 items, 4 responses for 7 items, and 1 response for 6 items.

\subsubsection{Annotation}
Across questionnaires, neutral labels dominated. Overall, 51.9\% of assigned labels were neutral, 28.6\% positive, and 19.6\% negative.
At the questionnaire level (30 statements), the median label counts were 16 neutral, 8 positive, and 6 negative.
Across rounds, these medians were very similar (neutral: 15--16; positive: 8--9; negative: 5--6), indicating a strongly imbalanced but largely stable label distribution over time.

\subsubsection{Intragroup Conflict Scale}
Task and relationship conflicts were assessed using the Intragroup Conflict Scale~\cite{lehmann2011intragroup}.
Overall, task conflicts had a median of 8, while relationship conflicts had a median of 5 (round-specific medians for relationship conflicts were 5--6).
Both measures were consistently below half of the maximum possible 24 points, indicating generally low conflict levels in the projects.
Scores ranged from the minimum of 4 to maxima of 22 (relationship) and 19 (task).

\subsection{Statistical Data Analysis}
Next, we tested the variables against the hypotheses outlined in Table~\ref{tab:merged_hypotheses}. We report uncorrected $p$-values for transparency and additionally report global multiple-testing corrections across all Spearman correlation tests ($m=29$) using Holm (FWER) and Benjamini--Hochberg (FDR). Correlations that reach $p<0.05$ but do not survive global correction are interpreted as exploratory trends.

We began by characterizing within-participant stability in sentiment labeling (RQ1) using Fleiss' $\kappa$ and by testing whether $\kappa(p)$ is associated with psychological factors (H$1_0$).
Table~\ref{tab:participants} summarizes participation counts and the resulting longitudinal subsamples.

First, it reports the number of submitted questionnaires in each survey round.
Second, it reports how many distinct participants completed at least $k$ rounds, which is the relevant sample size for analyses that require repeated measurements.
Third, it reports how many distinct participants completed exactly $k$ rounds, which makes the dropout pattern explicit.

\begin{table}[htb]
\caption{Participation counts per survey round and repetition frequency. The table distinguishes between (i) submitted questionnaires per round, (ii) distinct participants who took part in at least $k$ rounds, and (iii) distinct participants who took part in exactly $k$ rounds.}
\label{tab:participants}
\begin{tabularx}{\columnwidth}{lXXXX}
\toprule
\textbf{Survey rounds} & \textbf{1} & \textbf{2} & \textbf{3} & \textbf{4} \\
\midrule
Responses in round ($n$) & 55 & 31 & 27 & 33 \\
\midrule
\textbf{Participants with at least $k$ rounds} & $\boldsymbol{\geq 1}$ & $\boldsymbol{\geq 2}$ & $\boldsymbol{\geq 3}$ & $\boldsymbol{\geq 4}$ \\
\midrule
Distinct participants ($n$) & 81 & 33 & 21 & 11 \\
\midrule
\textbf{Participants with exactly $k$ rounds} & $\boldsymbol{= 1}$ & $\boldsymbol{= 2}$ & $\boldsymbol{= 3}$ & $\boldsymbol{= 4}$ \\
\midrule
Distinct participants ($n$) & 48 & 12 & 10 & 11 \\
\bottomrule
\end{tabularx}
\end{table}

To quantify within-participant self-agreement on label assignments, we computed Fleiss' $\kappa$ for participants with repeated measurements. Table~\ref{tab:kappas} reports the distribution of $\kappa$ for participants with at least two, at least three, and four completed rounds. With mean Fleiss' $\kappa$ values between 0.48 and 0.52, students showed only moderate agreement with themselves.

\begin{table}[htb]
\caption{Overview of Fleiss' $\kappa$ values (participant level).}
\label{tab:kappas}
\begin{tabularx}{\columnwidth}{Xrrrrrr}
\toprule
\textbf{Participation frequency} & \textbf{n}  & \textbf{$\mu$} & \textbf{Median} & \textbf{$\sigma$} & \textbf{min} & \textbf{max} \\ \midrule
At least two times   & 33 & 0.48 & 0.53 & 0.23 & -0.03 & 0.78 \\
At least three times & 21 & 0.52 & 0.55 & 0.18 &  0.08 & 0.78 \\
Four times           & 11 & 0.51 & 0.55 & 0.19 &  0.09 & 0.78 \\
\bottomrule
\end{tabularx}
\end{table}

\vspace{3pt}
\setlength{\shadowsize}{2pt}
\noindent
\shadowbox{
\begin{minipage}{0.94\columnwidth}
	\textbf{Observation:} We observe notable changes in the perceived polarity of statements across repeated submissions. That is, the same developer may label the same message differently at different project phases. 
\end{minipage}
}

Table~\ref{tab:corr_global} shows the correlations that reached the conventional uncorrected threshold of $p<0.05$.
For RQ1, we tested whether within-participant self-agreement in sentiment labeling ($\kappa(p)$) is associated with psychological factors (H$1.1_0$--H$1.5_0$, mood trait $mo$, reactivity $rea$, positive affect $pos$, negative affect $neg$, and life circumstances $liv$, cf.\ Table~\ref{tab:merged_hypotheses}). 
None of these correlations reached the uncorrected threshold of $p<0.05$.
Accordingly, we do not reject H$1_0$ and interpret the results for RQ1 as descriptive evidence of intra-individual variability (cf.\ Section~\ref{sec:results}) without statistical evidence that $\kappa(p)$ is systematically related to the measured psychological factors in our sample.

\begin{table}[htb]
\caption{Spearman correlations reaching $p<0.05$ (uncorrected). We additionally report global multiplicity adjustments across all correlation tests ($m=29$): Holm (FWER) and Benjamini--Hochberg (FDR).}
\label{tab:corr_global}
\begin{tabularx}{\columnwidth}{lXXrllll}
\toprule
\textbf{Hypo} & \multicolumn{2}{l}{\textbf{Correlation}} & \textbf{$n$} & $\boldsymbol{\varrho}$ & $\boldsymbol{p}$ & $\boldsymbol{p_{\mathrm{Holm}}}$ & $\boldsymbol{q_{\mathrm{BH}}}$ \\
\midrule
H$2.1_0$ & General mood ($mo$) & Positive labels ($num_{pos}$) & 128 & $\phantom{-}0.211$ & 0.017 & 0.469 & 0.243 \\
H$5.3_0$ & Negative affect ($neg$) & Negative labels ($num_{neg}$) & 146 & $\phantom{-}0.169$ & 0.041 & 1.000 & 0.298 \\
H$6.3_0$ & Life circumstances ($liv$) & Negative labels ($num_{neg}$) & 146 & $-0.205$ & 0.013 & 0.374 & 0.243 \\
H$8.3_0$ & Task conflicts ($tas$) & Negative labels ($num_{neg}$) & 146 & $\phantom{-}0.175$ & 0.035 & 0.943 & 0.298 \\
\bottomrule
\end{tabularx}
\end{table}

Given the large number of correlation tests, we additionally applied global multiple-testing corrections across all Spearman correlations ($m=29$).
Table~\ref{tab:corr_global} lists the correlations that reach the conventional uncorrected threshold ($p<0.05$) together with Holm-adjusted $p$ values (FWER) and Benjamini--Hochberg adjusted $q$ values (FDR).
None of the observed correlations remains statistically significant after global Holm or BH correction.
We therefore interpret these associations as exploratory trends rather than confirmatory evidence.

\vspace{3pt}
\setlength{\shadowsize}{2pt}
\noindent
\shadowbox{
\begin{minipage}{0.94\columnwidth}
\textbf{Observation:} Several small correlations reach $p<0.05$ before correction (e.g., life circumstances and negative labeling), but none remains significant after global multiplicity correction (Holm or BH-FDR). Accordingly, we treat these results as exploratory trends.
\end{minipage}
}

\subsubsection{Attrition and Dropout Bias}

Participation was uneven across survey rounds (Round~1: $n=55$, Round~2: $n=31$, Round~3: $n=27$, Round~4: $n=33$), resulting in an unbalanced longitudinal dataset with a total of 146 completed questionnaires from 81 distinct participants.
In terms of retention, 48 participants completed exactly one round, 12 completed exactly two rounds, 10 completed exactly three rounds, and 11 completed all four rounds.

\begin{table}[htb]
\caption{Retention distribution and baseline phase. Baseline phase is defined as the first submission per participant.}
\label{tab:attrition_phase}
\begin{tabularx}{\columnwidth}{lrrrr}
\toprule
\textbf{Baseline phase} & \textbf{1 round} & \textbf{2 rounds} & \textbf{3 rounds} & \textbf{4 rounds} \\
\midrule
Audit  & 7  & 0 & 0 & 0 \\
Break  & 9  & 2 & 0 & 0 \\
Iter1  & 24 & 7 & 8 & 10 \\
Iter2  & 8  & 3 & 2 & 1 \\
\bottomrule
\end{tabularx}
\end{table}

To assess potential dropout bias, we compared participants who completed exactly one round to those who completed at least two rounds.
First, the availability of mood-trait data (with\_mood: yes/no) was not associated with retention (Fisher's exact test: OR$=1.67$, $p=0.544$).

Second, baseline phase showed a structural association with retention: participants whose baseline occurred later in the project had fewer opportunities to complete additional rounds (Table~\ref{tab:attrition_phase}).
A Fisher test comparing baseline in Iteration~1 vs.\ later baselines indicated a significant association with completing at least two rounds ($p=0.0227$).

Baseline comparisons were then conducted on participants with baseline mood-trait data available (with\_mood=yes; $n=68$ baselines).
After Holm correction across all tested baseline variables, only one robust difference emerged:
participants who completed exactly one round assigned more negative labels at baseline than those who completed at least two rounds (median $num_{neg}=8$ vs.\ 5). Mann--Whitney $p=0.00230$, Holm-$p=0.0252$, and Cliff's $\delta=0.433$ indicate a medium effect.
All other baseline variables (mood trait, reactivity, PANAS positive/negative affect, life circumstances, relationship/task conflicts, and English proficiency) showed no robust differences after Holm correction.

As a sensitivity analysis, restricting baselines to Iteration~1 ($n=49$ baselines) replicated the pattern (median 8 vs.\ 5). Mann--Whitney $p=0.00147$, Holm-$p=0.0162$, and Cliff's $\delta=0.528$ indicate a medium effect.
Logistic regression further confirmed that baseline negativity predicted lower retention:
for Iteration~1 baselines, each additional negative label at baseline was associated with lower odds of completing at least two rounds (OR$=0.659$, 95\% CI [0.496, 0.874], $p=0.00382$).
Using all baselines (with\_mood=yes) and controlling for whether the baseline occurred in Iteration~1, the effect remained (baseline $num_{neg}$: OR$=0.695$, 95\% CI [0.551, 0.876], $p=0.00205$). Baseline Iteration~1 also had an effect (OR$=6.35$, $p=0.0128$).

\paragraph{Robustness: retention model and IPW weights.}
As an additional robustness check, we fitted a participant-level retention model using baseline psychological predictors ($n=65$ complete cases) and derived stabilized inverse-probability weights (IPW).
None of the predictors was significantly associated with retention (all $p \ge 0.369$), and weights were moderate (mean 0.995; max 1.613), suggesting that weighting would not materially change conclusions.

\vspace{3pt}
\setlength{\shadowsize}{2pt}
\noindent
\shadowbox{
\begin{minipage}{0.94\columnwidth}
\textbf{Observation:} Attrition is partly structural (baseline phase limits opportunity for repeated participation) and is not explained by baseline mood, affect, life circumstances, or conflict measures. However, one-time participants assign more \emph{negative} sentiment labels at baseline than participants who complete multiple rounds. IPW weights are moderate (max 1.613), suggesting weighting would not materially alter conclusions.
\end{minipage}
}

\subsubsection{Statement-level multivariate analysis (GEE)}
To leverage all available statement-level observations while accounting for repeated measures, we estimated logistic GEE models on the subset with mood-trait data available (with\_mood=yes).
This subset comprises 128 questionnaires from 68 participants, resulting in 3840 statement labels (30 labels per questionnaire).
Label distributions were imbalanced: 1089 positive labels (28.4\%), 2013 neutral labels (52.4\%), and 738 negative labels (19.2\%).

\begin{table*}[htbp]
\caption{Statement-level GEE results (one-vs-rest logistic models).
Reported values are odds ratios (OR) with 95\% confidence intervals in brackets and corresponding $p$-values.
Odds ratios correspond to a one-standard-deviation increase in the predictor (z-standardized).
Models include phase and statement fixed effects. Standard errors are clustered by participant.
Subset: with\_mood=yes (128 questionnaires, 68 participants, 3840 labels).}
\label{tab:gee_statementlevel}
\begin{tabularx}{\textwidth}{lXXX}
\toprule
\textbf{Predictor (z)} &
\textbf{Positive vs.\ not} (OR [95\% CI], $p$) &
\textbf{Negative vs.\ not} (OR [95\% CI], $p$) &
\textbf{Neutral vs.\ not} (OR [95\% CI], $p$) \\
\midrule
$mo$            & 1.45 [1.07, 1.96], 0.017 & 1.16 [0.95, 1.42], 0.141 & 0.82 [0.68, 0.98], 0.028 \\
$rea$           & 1.28 [1.09, 1.51], 0.002 & 0.97 [0.84, 1.12], 0.672 & 0.91 [0.78, 1.07], 0.262 \\
$pos$ (PANAS)   & 1.03 [0.85, 1.24], 0.760 & 0.93 [0.81, 1.07], 0.329 & 1.10 [0.93, 1.29], 0.257 \\
$neg$ (PANAS)   & 0.97 [0.82, 1.13], 0.680 & 1.11 [0.97, 1.28], 0.126 & 0.93 [0.83, 1.05], 0.231 \\
$liv$           & 0.96 [0.82, 1.12], 0.585 & 0.91 [0.77, 1.08], 0.286 & 0.99 [0.83, 1.19], 0.949 \\
$tas$           & 1.03 [0.87, 1.21], 0.766 & 1.12 [0.99, 1.27], 0.081 & 0.96 [0.87, 1.06], 0.395 \\
$rel$           & 1.00 [0.84, 1.21], 0.964 & 0.96 [0.82, 1.12], 0.603 & 0.99 [0.86, 1.15], 0.895 \\
$eng$           & 1.21 [1.03, 1.43], 0.023 & 1.18 [0.99, 1.41], 0.063 & 0.93 [0.84, 1.04], 0.218 \\
$idx(p,r)$      & 0.97 [0.79, 1.19], 0.759 & 1.04 [0.85, 1.27], 0.697 & 1.07 [0.90, 1.26], 0.443 \\
\bottomrule
\end{tabularx}
\end{table*}

Table~\ref{tab:gee_statementlevel} summarizes odds ratios (per one standard deviation increase in the predictor) from three one-vs-rest models (positive, negative, neutral).
Controlling for statement fixed effects and phase, higher mood trait is associated with higher odds of assigning a \emph{positive} label (OR=1.45, 95\% CI [1.07, 1.96], $p=0.017$) and lower odds of assigning a \emph{neutral} label (OR=0.82, 95\% CI [0.68, 0.98], $p=0.028$).
Higher reactivity is also associated with higher odds of assigning a positive label (OR=1.28, 95\% CI [1.09, 1.51], $p=0.002$).
In addition, higher self-reported English proficiency is associated with more positive labeling (OR=1.21, 95\% CI [1.03, 1.43], $p=0.023$).

For the \emph{negative} label model, task-related conflict shows a small positive trend (OR=1.12, 95\% CI [0.99, 1.27], $p=0.081$), and English proficiency shows a similar trend (OR=1.18, 95\% CI [0.99, 1.41], $p=0.063$), while other predictors show no clear association.
Across all three models, phase fixed effects and participation index effects were not statistically significant, suggesting no strong overall shift in labeling attributable to project phase order or repeated participation when controlling for statement-specific differences.

\vspace{3pt}
\setlength{\shadowsize}{2pt}
\noindent
\shadowbox{
\begin{minipage}{0.94\columnwidth}
\textbf{Observation:} In statement-level GEE models controlling for statement- and phase-specific differences, higher mood trait, higher reactivity, and higher English proficiency are associated with more \emph{positive} sentiment assignments. Mood trait is also associated with fewer \emph{neutral} assignments. Negative labeling shows at most weak trends, for example with task-related conflict, indicating that positivity-related effects are more pronounced than negativity-related effects in this dataset.
\end{minipage}
}

\subsubsection{Statement-level disagreement and within-participant flipping}

Table~\ref{tab:statement_metrics_summary} provides an overview of our statement-level heterogeneity metrics across the 30 statements, summarizing overall disagreement across submissions (entropy and majority share) and within-participant instability over time (flip rate and strong flip rate for participants with $\geq 2$ rounds, $n=33$).

\begin{table}[htb]
\caption{Summary of statement-level disagreement and within-participant flipping (30 statements).}
\label{tab:statement_metrics_summary}
\begin{tabularx}{\columnwidth}{lrrr}
\toprule
\textbf{Metric} & \textbf{Mean} & \textbf{Min} & \textbf{Max} \\
\midrule
Entropy ($H$) & 1.107 & 0.633 & 1.539 \\
Majority share & 0.676 & 0.445 & 0.884 \\
Flip rate (participants $\geq 2$ rounds, $n=33$) & 0.365 & 0.121 & 0.606 \\
Strong flip rate (pos$\leftrightarrow$neg) & 0.037 & 0.000 & 0.121 \\
\bottomrule
\end{tabularx}
\end{table}

Across all submitted questionnaires (81 participants, 146 questionnaires), participants provided 4380 sentiment labels for the 30 statements (146 labels per statement, aggregated across submissions).
Because participants could contribute multiple questionnaires, these 146 labels per statement do not represent 146 independent annotators. Instead, participants with more submissions contribute more labels.

We observed substantial statement-level heterogeneity.
On average, statement entropy was 1.107 (min 0.633, max 1.539), and the average majority share was 0.676 (min 0.445, max 0.884), indicating that some statements attracted broad consensus whereas others were perceived very inconsistently.

To assess whether statements with mixed overall labeling across submissions also tend to be unstable within participants, we analyzed within-person label changes for participants with at least two rounds ($n=33$).
Across statements, the average flip rate was 0.365 (min 0.121, max 0.606), meaning that for an average statement, 36.5\% of participants changed their label at least once across rounds.
Strong flips (pos$\leftrightarrow$neg) were comparatively rare (mean 0.037; max 0.121), suggesting that most changes involve the neutral category rather than direct polarity reversals.

Statement entropy and flip rate were strongly positively correlated ($r=0.823$), indicating that statements with more mixed overall labeling across submissions also tend to be the ones for which individuals are more likely to change their perception over time.
Table~\ref{tab:statement_entropy_top} lists examples of the most disagreement-prone statements.

\begin{table*}[htbp]
\caption{Statements with the highest disagreement (top-6 by entropy). Counts are aggregated over all questionnaires (i.e., across submissions. $n=146$ labels per statement). Flip rates are computed over participants with $\geq 2$ rounds ($n=33$).}
\label{tab:statement_entropy_top}
\begin{tabularx}{\textwidth}{rXlrrrrrr}
\toprule
\textbf{ID} & \textbf{Source} & \textbf{GT} & \textbf{pos} & \textbf{neu} & \textbf{neg} & \textbf{Maj.\ share} & \textbf{Entropy} & \textbf{Flip rate} \\
\midrule
9  & GitHub         & neg & 35 & 65 & 46 & 0.445 & 1.539 & 0.545 \\
21 & Stack Overflow & pos & 40 & 72 & 34 & 0.493 & 1.504 & 0.606 \\
27 & GitHub         & pos & 78 & 39 & 29 & 0.534 & 1.455 & 0.333 \\
6  & GitHub         & pos & 16 & 63 & 67 & 0.459 & 1.388 & 0.364 \\
8  & GitHub         & neg & 21 & 84 & 41 & 0.575 & 1.376 & 0.515 \\
24 & Stack Overflow & neu & 16 & 83 & 47 & 0.568 & 1.339 & 0.455 \\
\bottomrule
\end{tabularx}
\end{table*}

\vspace{3pt}
\setlength{\shadowsize}{2pt}
\noindent
\shadowbox{
\begin{minipage}{0.94\columnwidth}
\textbf{Observation:} Statements differ strongly in how consistently they are perceived. Statements with more mixed overall labeling across submissions (higher entropy / lower majority share) are also more likely to trigger within-participant label changes over time (flip rate), suggesting substantial statement-driven variability in sentiment perception.
\end{minipage}
}

\subsubsection{Labeling rationale and uncertainty meta-data}

\begin{table}[htb]
\caption{Labeling rationale and uncertainty meta-data (questionnaire level, $n=146$).}
\label{tab:label_meta_reasons}
\begin{tabularx}{\columnwidth}{Xrr}
\toprule
\textbf{Item} & \textbf{n} & \textbf{\%} \\
\midrule
\textbf{Criteria used for labeling} & & \\
Content & 112 & 76.7 \\
Tone & 105 & 71.9 \\
Emoticons & 69 & 47.3 \\
Other (free text provided) & 13 & 8.9 \\
\midrule
\textbf{Reasons for uncertainty} & & \\
Too many possible interpretations & 81 & 55.5 \\
Contradictory emotions & 53 & 36.3 \\
No label fit well & 45 & 30.8 \\
No insecurities reported & 30 & 20.5 \\
Other (free text provided) & 9 & 6.2 \\
\bottomrule
\end{tabularx}
\end{table}

Across all questionnaires ($n=146$), participants reported which criteria they used for assigning sentiment labels.
Most participants relied on statement \emph{content} (112/146, 76.7\%) and perceived \emph{tone} (105/146, 71.9\%).
Nearly half reported considering \emph{emoticons} (69/146, 47.3\%).
A minority provided additional free-text explanations (13/146, 8.9\%).

Participants also indicated sources of uncertainty in their labeling.
The most frequent reason was that there were \emph{too many possible interpretations} (81/146, 55.5\%).
Further reasons were \emph{contradictory emotions} (53/146, 36.3\%) and that \emph{no label fit well} (45/146, 30.8\%).
A subset reported \emph{no insecurities} (30/146, 20.5\%). Optional free-text explanations for uncertainty were provided in 9/146 questionnaires (6.2\%).
Table~\ref{tab:label_meta_reasons} summarizes these distributions.

Overall confidence in labeling (1=very unsure, 5=very sure) had a mean of 3.43 (median 4, IQR 3--4), indicating moderate to high self-reported confidence.
Confidence showed no clear association with the participant's mismatch rate against reference labels ($\rho=-0.143$, $p=0.0848$) nor with the entropy of the participant's own label distribution ($\rho=0.033$, $p=0.690$).

However, confidence differed systematically by reported uncertainty flags.
Participants who reported \emph{no insecurities} had higher confidence (median 4 vs.\ 3). Mann--Whitney $p=3.82\times10^{-7}$, Holm-$p=3\times10^{-6}$, and BH-$q=3\times10^{-6}$.
Conversely, participants who reported \emph{too many interpretations} had lower confidence (median 3 vs.\ 4). $p=0.00207$, Holm-$p=0.0166$, and BH-$q=0.00932$.
Other uncertainty flags did not show robust differences after correction.

\vspace{3pt}
\setlength{\shadowsize}{2pt}
\noindent
\shadowbox{
\begin{minipage}{0.94\columnwidth}
\textbf{Observation:} Most participants base sentiment labels on content and tone, and more than half report uncertainty due to multiple plausible interpretations. Confidence aligns strongly with perceived uncertainty (especially ``no insecurities'' vs.\ ``too many interpretations''), but does not reliably predict disagreement with reference labels.
\end{minipage}
}

\section{Discussion}
\label{sec:discussion}

\subsection{Answering the Research Questions}
Based on our results presented in Section~\ref{sec:results}, we address the research questions as follows:

\vspace{6pt}\noindent\textbf{RQ1:} \textit{How stable is the perception of individual developers towards textual statements?}
\newline\noindent
We found notable intra-individual variability in sentiment perception over time. Across participants with repeated measurements, self-agreement ranged from low to high (Fleiss’ $\kappa$ between $-0.03$ and $0.78$), indicating substantial differences in self-consistency. In addition, statement-level analyses show that this variability is not uniform across items. Statements differ strongly in overall disagreement across submissions (mean entropy 1.107; mean majority share 0.676), and among participants with repeated participation ($n=33$), the average flip rate per statement is 0.365. Statements with more mixed overall labeling across submissions (higher entropy / lower majority share) were also more likely to ``flip'' within the same person over time ($r=0.823$). Overall, sentiment perception is not fully stable even within individuals and is partly driven by statement-specific ambiguity.

\vspace{6pt}\noindent\textbf{RQ2:} \textit{How are team-internal (psychological) aspects such as mood or conflicts related to how developers perceive text-based communication?}
\newline\noindent
In the correlation battery, several associations reached $p<0.05$ before correction, but none remained significant after global multiple-testing correction (Holm or Benjamini--Hochberg). We therefore treat these as exploratory trends. To better account for repeated measures and statement heterogeneity, we additionally conducted statement-level multivariate analyses using logistic GEE models (clustered by participant), including statement fixed effects and controlling for project phase, participation index (submission order), and English proficiency. These models indicate that higher mood trait and higher reactivity are associated with higher odds of assigning \emph{positive} labels, and mood trait is also associated with lower odds of assigning \emph{neutral} labels. Effects related to negative labeling are weaker. Task conflict shows at most a small trend toward more negative labels. Overall, psychological factors appear to relate more consistently to \emph{positive} vs.\ \emph{neutral} sentiment assignments than to negative labeling, and observed effects are small.

\vspace{6pt}\noindent\textbf{RQ3:} \textit{How are external factors such as project phases related to how developers perceive text-based communication?}
\newline\noindent
We examined project-phase effects in two ways. First, correlation analyses between project phase order and labeling counts showed no evidence of phase-related shifts in sentiment assignments ($phase_{ord}$ vs.\ $num_{pos}$: $\rho=-0.07$, $p=0.385$; vs.\ $num_{neu}$: $\rho=0.10$, $p=0.244$; vs.\ $num_{neg}$: $\rho=-0.12$, $p=0.154$). Second, in statement-level GEE models that controlled for statement fixed effects and clustered repeated measures within participants, phase effects were not statistically significant. Thus, within the project phases captured in our survey, we find no clear evidence that project timing systematically changes how statements are perceived. Any phase-related influence, if present, may be smaller than statement-specific ambiguity and within-person variability, or may require more direct measures of phase-related stress and contextualized project communication to detect.

\subsection{Interpretation}

A central result of our study is that sentiment perception variability is strongly statement-driven.
Across the 30 statements, we observed substantial heterogeneity in how consistently they were perceived. Statement entropy was high on average (mean 1.107, up to 1.539). The majority share ranged widely (0.445--0.884).
Importantly, statements that showed more mixed labeling across submissions were also the ones that were more likely to flip within the same participant over time. We found a strong association between entropy and flip rate ($r=0.823$).
This suggests that intra-individual instability is not merely random noise or a general tendency of a person. Instead, it is concentrated on specific, ambiguity-prone messages, such as short or context-dependent utterances and statements that can plausibly be read as neutral or affective depending on interpretation.

\textit{Mood and psychological factors: correlations vs.\ multivariate models.}
In the correlation battery, several associations reached $p<0.05$ before correction. None remained significant after global multiplicity adjustment using Holm or Benjamini--Hochberg correction.
Accordingly, we interpret the correlation-level signals as exploratory rather than confirmatory evidence.
At the same time, statement-level multivariate GEE models account for repeated measures by clustering within participants. They also control for statement heterogeneity using statement fixed effects. These models indicate small but systematic associations.
Higher mood trait and higher reactivity are associated with higher odds of assigning positive labels. Mood trait is also associated with lower odds of neutral labeling.
In contrast, predictors of negative labeling are weaker. Task conflict shows at most a small trend toward more negative assignments.
Taken together, the data suggest that mood-related factors, if they matter, may primarily shift borderline statements from neutral toward positive rather than driving outright negative interpretations.

\textit{Attrition and a potential negativity-related retention bias.}
Our attrition analysis indicates that retention is partly structural. Baseline phase limits the opportunity for repeated participation.
Baseline mood, affect, life circumstances, and conflict measures do not robustly explain dropout.
However, one robust baseline difference emerged. Participants who completed only one survey round assigned more negative labels at baseline than participants who completed multiple rounds. The effect was of medium magnitude and it replicated in an Iteration~1-only sensitivity analysis.
This pattern implies a potential negativity-related retention bias in the longitudinal subsample. It may attenuate or distort observed relationships involving negative labels when analyses rely more heavily on repeat participants.
Practically, this finding underscores that dropout in longitudinal surveys may not be random with respect to labeling behavior, even if it appears largely unrelated to standard psychological covariates.

\textit{Labeling rationale and uncertainty as evidence for contextual ambiguity.}
Participants’ self-reported labeling criteria further support the role of interpretive ambiguity.
Most participants relied on content and tone. Nearly half considered emoticons. More than half reported uncertainty due to multiple plausible interpretations.
Confidence aligned strongly with self-reported uncertainty flags, such as no insecurities versus too many interpretations. Confidence did not clearly predict disagreement with reference labels.
Free-text responses were sparse. When present, they most frequently pointed to missing context. This matches the observed statement-level disagreement and flipping patterns.
Overall, these meta-data suggest that the core challenge in sentiment perception is not only the individual's state. It also reflects limited contextual cues and interpretive flexibility of brief, decontextualized statements.

\textit{Comparison with Industry Projects.}
A frequent concern in empirical software engineering research is the use of student developers as participants. Student projects differ from professional environments in incentives, experience, team maturity, and consequences of failures. These differences may affect communication and interpretation.
Prior work suggests that students and professionals can behave similarly in some controlled settings, particularly when both groups are unfamiliar with the evaluated technique, such as test-driven development~\cite{salman2015studentindustry,pankaj2019softwareuniindustry}. However, evidence on transferability is mixed.
In particular, Counsell et al.~\cite{Counsell2008studentindustry} highlight that even when aggregate outcomes appear similar, the underlying judgment strategies can differ between students and industry participants.
Accordingly, we interpret our results primarily in the context of software engineering education and discuss implications for professional practice cautiously.

In line with this perspective, we argue that the observed combination of strong statement-driven heterogeneity and within-person variability concentrated on ambiguity-prone messages may also be relevant in professional settings. Short text snippets, such as chat messages and issue comments, are frequently read without rich context.
However, the extent to which these patterns transfer to in-project communication with shared history remains an open question.

\textit{Practical Implications for Teams.}
Our findings emphasize that sentiment perception is context-sensitive and can vary even within the same person, particularly for messages that are inherently ambiguous.
For practitioners, this suggests establishing communication practices that reduce interpretive flexibility. Examples include adding minimal context, stating intent explicitly for short messages, and avoiding sarcasm or irony in critical situations.
Regular feedback loops, such as retrospectives or short check-ins, can help detect mismatched perceptions early.

The small trend that task conflict coincides with more negative labeling suggests that conflict management may matter not only for coordination but also for how messages are interpreted.
Even task-focused disagreements may color sentiment perception, especially for borderline statements that can be read as neutral or affective.

\textit{Perception Changes Within Individuals.}
The observed within-person variability and strong statement-driven effects suggest that automated sentiment analysis systems should be applied with caution.
Tools that assume stable interpretation across users or time may mischaracterize team climate, particularly when monitoring short, decontextualized utterances.
Incorporating statement-level uncertainty, such as flagging high-disagreement statements, and combining automated predictions with lightweight human feedback may improve both accuracy and acceptance.

\subsection{Threats to Validity}
\label{sec:threats}
We discuss potential limitations of our study along the categories defined by Wohlin et al.~\cite{wohlin2012experimentation}: construct, internal, external, and conclusion validity.

\textit{Construct Validity.}
We used established, peer-reviewed instruments to measure mood traits~\cite{bohner1991stimmungs}, affect using PANAS~\cite{watson1988development,krohne1996untersuchungen}, and group conflicts~\cite{lehmann2011intragroup}, which supports construct validity.
Nevertheless, while these scales have been widely validated, they were not developed specifically for software engineering contexts and may not capture all nuances relevant to developer behavior.
The labeling-rationale and uncertainty items (criteria used, uncertainty reasons, and confidence) were designed for this study and were not externally validated instruments. Thus, measurement precision for these meta-data may be limited.

A further limitation is the use of isolated, pre-existing text snippets from GitHub and Stack Overflow.
These lack contextual cues such as project history or prior interactions, which limits ecological validity.
However, this choice allowed us to standardize the labeling task across participants.
Our statement-level analyses confirm that some snippets are intrinsically ambiguous:
they show more mixed overall labeling across submissions and higher within-participant label changes over time, which indicates that a substantial portion of the observed variability is driven by statement characteristics rather than only by participant state.
This supports the need for modeling statement heterogeneity explicitly and suggests that context-rich communication may yield different patterns.

\textit{Internal Validity.}
As with any survey-based research, biases such as misinterpretation, fatigue, or social desirability may influence responses.
We mitigated these risks by using validated instruments, randomized statement order, and clear instructions.
We excluded incomplete questionnaires.
In a small number of cases, participants submitted the questionnaire more than once within the same survey round. We retained complete submissions and model repeated participation explicitly (participation index), including it as a control in the statement-level models.

Dropout effects remain a threat because participation decreased across rounds.
We explicitly assessed dropout bias by comparing baseline characteristics of participants with exactly one completed round to those with at least two rounds.
We found that retention is partly structural because baseline phase determines how many further rounds were available.
Most baseline psychological measures did not robustly differ between retention groups after correction for multiple comparisons.
However, one-time participants assigned more negative labels at baseline, indicating a potential negativity-related retention bias in the longitudinal subsample.
We therefore interpret results involving negative labeling, and analyses that rely primarily on repeat participants, with additional caution.

We also did not capture team membership and team size in a way that supports consistent modeling of team-level clustering or covariates.
If team-specific norms or dynamics influence sentiment perception, the inability to control for team effects could confound some associations.

\textit{Conclusion Validity.}
Our study is exploratory and involves many potential associations.
To reduce false positives, we applied global multiple-testing correction across the full correlation battery, reporting both Holm-adjusted $p$ values (FWER) and Benjamini--Hochberg adjusted $q$ values (FDR).
Under global correction, none of the correlation-level effects remains statistically significant.
We therefore frame correlation results as exploratory trends.

We also strengthened inference by adding multivariate statement-level models using generalized estimating equations (GEE).
These models account for repeated measures by clustering within participants and control for statement heterogeneity via statement fixed effects.
This reduces the risk that results are artifacts of non-independence or a few highly ambiguous statements.
Nevertheless, effect sizes remain small, and some analyses rely on subsets due to missing covariates.
In addition, label imbalance, with neutral labels dominating, limits power for detecting predictors of negative labeling and can widen uncertainty around estimates.

\textit{External Validity.}
A key limitation is that the study was conducted with students rather than professional developers.
Differences in incentives, experience, and communication maturity may affect both mood dynamics and how messages are interpreted.
Prior work suggests that students and professionals can show similar behavior in some controlled or technique-focused tasks, such as when adopting unfamiliar development approaches like TDD~\cite{salman2015studentindustry,pankaj2019softwareuniindustry}.
At the same time, transferability to professional practice is not guaranteed.
Counsell et al.~\cite{Counsell2008studentindustry} report no significant differences in cohesion ratings in their setting, while also noting differences in how students and industry participants formed these judgments.
We therefore refrain from strong generalizations beyond the educational context.

Our findings should also be interpreted in light of the decontextualized stimuli.
The labeling meta-data indicate that many participants experienced uncertainty due to multiple plausible interpretations and missing context.
This may amplify ambiguity effects relative to real team communication where shared history and conversational context exist.
Future studies should replicate the analyses using contextualized in-project communication.

The inclusion of a holiday break is another contextual difference.
However, we found no evidence that project phase order is associated with systematic shifts in labeling distributions.
Finally, while most participants shared a similar cultural background (German-speaking), we did not analyze cultural or linguistic factors in depth.
Since these may influence sentiment interpretation, broader studies should include more diverse participant groups and directly model language proficiency and cultural context.

\subsection{Future Work}
\label{sec:future}

Identifying the factors that shape sentiment perception in software development remains a complex challenge. This study offers an initial exploration but also highlights several areas where further research is needed to validate and expand the findings.

First, future work should replicate our study in professional software development environments to assess whether the observed patterns hold under real-world conditions. This includes within-person variability as well as the potential role of task conflicts. Replication with experienced developers would strengthen external validity, especially regarding organizational context, stressors, and communication complexity~\cite{salman2015studentindustry,pankaj2019softwareuniindustry}.

Second, future studies should explicitly model and manipulate context. Our statement-level analyses indicate strong heterogeneity between statements and show that ambiguous snippets drive both disagreement between people and label changes within the same person over time. Follow-up work should therefore include richer, contextualized stimuli such as message threads, prior turns, or known team history. Experimental variants could systematically add or remove context to quantify how much context reduces ambiguity and perception flips.

Third, future work should explicitly capture and model team-level factors. Collecting team identifiers, team size, roles, and prior familiarity would enable testing whether sentiment perception varies systematically by team and whether team composition moderates the relationship between psychological factors and labeling behavior. With richer team data, future studies could incorporate team-level effects (e.g., team fixed/random effects) alongside participant- and statement-level variability.

Fourth, a notable limitation of our study is the use of English-language statements, despite participants primarily communicating in German. Follow-up studies should investigate the impact of language proficiency and cultural background on sentiment perception. Emotional tone, idioms, and sarcasm can be interpreted differently by non-native speakers, potentially affecting how sentiment is perceived and labeled. Future designs should measure language proficiency more granularly and include statements in participants' primary language.

Fifth, future studies should broaden the scope of conflict types considered. We focused on task and relationship conflicts, but process-related tensions, technical debates, or disagreements about conventions may have distinct effects. A more event-based design that links perception to concrete episodes, such as deadline pressure or conflicts in specific tasks, could better capture how conflict and communication interact over time.

From a methodological perspective, future research should continue moving beyond pairwise correlations. Larger datasets would enable fully multilevel models that jointly account for participant- and statement-level random effects with multinomial outcomes, and allow testing interactions between psychological and contextual variables. Such models would also support stronger inferences about how much variance is attributable to person effects versus statement effects, which our results suggest is a key distinction.

Another challenge is retention in longitudinal studies. Our attrition analysis suggests that dropout is partly structural and may also be selective with respect to labeling behavior, since one-time participants labeled statements more negatively at baseline. Future work should therefore prioritize retention strategies and also plan analyses that are robust to selective dropout. Practical steps include shorter instruments, staged incentives, or embedded feedback that increases participants' motivation to continue.

Finally, increasing ecological validity remains an open goal. While we used decontextualized statements to ensure comparability, future work should incorporate real project communication and combine automated methods with qualitative approaches. In particular, collecting per-statement uncertainty ratings and short justifications during labeling would allow linking ambiguity, confidence, and perception changes more directly, and could yield more actionable insights for improving communication in both academic and industrial software teams.

\section{Conclusion}
\label{sec:conclusion}

Teamwork is a central component of modern software development. Prior research suggests that developers' affect and well-being can relate to problem solving and collaboration~\cite{graziotin2014happy}, making socio-emotional factors relevant in team settings. To assess emotional tone in communication, sentiment analysis tools are commonly used, yet their usefulness depends in part on how consistently individuals perceive sentiment in written messages. Prior work has shown variability in sentiment labeling, both between and within individuals~\cite{herrmannSentiSurvey22}.

To better understand what relates to sentiment perception, we conducted a multi-round survey study with 81 computer science students engaged in team-based software projects. Across four survey rounds, participants labeled 30 decontextualized statements (from GitHub and Stack Overflow) as positive, neutral, or negative and reported mood traits and states, life circumstances (computed as a mean score with \enquote{Not applicable} treated as missing), and perceived intragroup conflicts. Participants also provided meta-information about labeling rationale, uncertainty, and confidence.

Our results indicate notable intra-individual variability in sentiment perception over time. Self-agreement across repeated participation was only moderate, and statement-level analyses show that changes are concentrated on ambiguity-prone statements that also exhibit more mixed overall labeling across submissions. At the questionnaire level, a small number of bivariate correlations reached the conventional uncorrected threshold ($p<0.05$), but none remained statistically significant after global multiple-testing correction (Holm or Benjamini--Hochberg), and we therefore interpret them as exploratory trends. Complementing these results, statement-level repeated-measures models (logistic GEE with participant clustering and statement fixed effects) suggest small but systematic associations: higher mood trait and higher reactivity are associated with more positive labeling (and mood trait with less neutral labeling), while predictors of negative labeling are weaker and at most trend-level (e.g., task conflict). Across analyses, we find no clear evidence that project phase systematically shifts labeling when controlling for statement effects and repeated participation.

These findings highlight an important caveat for the use of sentiment analysis tools in software engineering. Such tools often implicitly assume that sentiment interpretation is stable or generalizable across people and time. Our results suggest that sentiment perception can fluctuate within the same individual and is strongly statement-dependent, especially for ambiguous, context-poor utterances. Accordingly, sentiment analysis should be applied cautiously when drawing conclusions about team climate or developer well-being, and should ideally be complemented with contextual information and mechanisms to capture uncertainty.

Our study was conducted in a software engineering education setting, and differences to professional environments should be expected. While some prior work suggests similarities between students and professionals in selected settings, transferability is not guaranteed. We therefore interpret our findings primarily within the educational context and view implications for professional practice as tentative indications, motivating replication with professional developers and contextualized in-project communication.

In conclusion, sentiment analysis can be a useful tool for understanding communication patterns in software teams. To leverage its potential responsibly, it should be combined with awareness of within-person variability, statement ambiguity, and selective attrition in longitudinal designs. Accounting for these factors can help foster clearer communication, reduce misunderstandings, and support healthier team dynamics.

\section*{Data Availability Statements}
The raw dataset containing the unprocessed survey results and associated metadata is available on \href{https://doi.org/10.5281/zenodo.18062116}{Zenodo}~\cite{obaidi2025swpDataset}.

\section*{Conflicts of Interest}
The authors declare no conflicts of interest.

\bibliography{references} 




\end{document}